\documentclass{article}
\usepackage{spconf,amsmath,amssymb,graphicx,booktabs,multirow}
\usepackage[table]{xcolor}
\usepackage{comment}
\usepackage{hyperref}

\newcommand{\OurMethod}{TIMUSE}

\title{\OurMethod: Text-Initialized Masked Semantic Token Modeling for Universal Speech Enhancement}

\name{Yiwei Guo$^{1,3}$, Hao Qiu$^{1}$, Mohan Ye$^1$, Zhijie Deng$^1$, Zehua Chen$^{2*}$, Kai Yu$^{1,3*}$}
\address{$^1$School of Computer Science, Shanghai Jiao Tong University\\$^2$Department of Computer Science and Technology, Tsinghua University\\$^3$Jiangsu Key Lab of Language Computing\thanks{$^*$Corresponding authors.}}

\begin{document}
\ninept

\maketitle

\begin{abstract}
Discrete-token-based universal speech enhancement (USE) enables strong sequence modeling but often suffers from a quality--fidelity trade-off, where perceptually plausible outputs may alter linguistic content or speaker characteristics. We present \OurMethod{}, a two-stage USE framework combining text-initialized masked semantic restoration with acoustic reconstruction. A pretrained diffusion language model (DLM) is adapted with LoRA to recover clean semantic tokens using masked generative modeling, while a conditional flow matching model reconstructs clean 24\,kHz speech from the recovered semantic tokens and original degraded speech.  Both system-level and controlled comparisons show that \OurMethod{} achieves a better balance between perceptual quality and fidelity than existing discrete USE models. Ablations further show that text DLM initialization brings a targeted gain under severe degradations.

\end{abstract}
\vspace{-0.05in}


\vspace{-0.11in}
\section{Introduction}
\vspace{-0.09in}
\label{sec:introduction}

Universal speech enhancement (USE) aims to recover clean speech from diverse
degradations, including noise, reverberation, bandwidth limitation, clipping,
and their mixtures~\cite{zhang2025aatc,li2026urgent}.
Severe degradations may not only corrupt speech but also remove linguistic
evidence, making generative USE methods increasingly popular.

Recent generative USE methods operate in either continuous or discrete
representation spaces.
Continuous-domain systems directly estimate waveforms, spectra, or acoustic
features~\cite{li2025bsrnnflow,rong2026unipase}, while discrete-token methods
reformulate enhancement as sequence generation.
The latter have shown strong perceptual quality and provide a natural interface
to language-model-based unified speech processing.
However, their generative flexibility also introduces a quality--fidelity
trade-off: perceptually plausible outputs may alter linguistic content or
speaker characteristics.
This problem is particularly important for USE, where reliable restoration is
required in addition to perceptual quality.

Existing discrete USE systems have explored different token representations and
generation strategies.
Acoustic-token methods~\cite{zhang2025anyenhance,liu2025udse,yan2026unise} directly generate codec representations.
Since linguistic and acoustic information are modeled jointly, this usually leads to heavier modeling burden.
Instead, semantic tokens provide a more compact representation of phonetic and
linguistic content~\cite{discretereview}.
Metis-SE~\cite{wang2025metis} explores this property by first restoring semantic tokens, followed by semantic-to-acoustic token generation.
SenSE~\cite{li2025sense} instead uses a flow matching model conditioned on the degraded
Mel spectrogram and restored semantic tokens, improving content fidelity while retaining
speaker and acoustic information from the observation.
These advances highlight the potential of using semantic tokens for USE.
Nevertheless, the quality--fidelity trade-off remains, motivating a more
effective generative model for the semantic restoration stage.

Recent progress in diffusion language models (DLMs) provides a promising solution.
In practice, mask-based DLMs are close to MaskGIT~\cite{chang2022maskgit}-style masked generative modeling (MGM) with diffusion-derived training criterion and inference strategies.
Current text DLMs have become efficient and competitive with similarly sized autoregressive (AR) language models while learning strong bidirectional sequence priors~\cite{ye2025dream,nie2026improved}.
These properties are well suited to the semantic restoration task in USE, where the complete
degraded utterance is available and missing content may need to be inferred
from bidirectional linguistic context.
Moreover, text pretraining provides a linguistic prior when acoustic evidence is severely degraded.
This motivates transferring a pretrained text DLM to semantic-token USE.
More broadly, discrete semantic tokens also provide a natural interface between
natural-language task specification and speech signal processing, making such
pretraining attractive for future unified enhancement, separation, extraction,
and editing systems.

Based on these observations, we present \OurMethod{}, a two-stage USE framework
combining text-initialized masked semantic restoration with
flow-matching-based acoustic reconstruction.
In the semantic restoration stage, we finetune a pretrained Dream~\cite{ye2025dream} DLM with LoRA~\cite{hu2022lora} to predict clean single-codebook $\mathcal{S}^3$ semantic tokens~\cite{du2024cosyvoice} from the encoded degraded speech.
Then, a conditional flow matching (CFM) model reconstructs the clean Mel spectrogram from the generated tokens and the original degraded-speech Mel spectrogram.
Figure \ref{fig:main} shows the workflow of \OurMethod{}.
On the URGENT 2026 benchmark~\cite{li2026urgent}, \OurMethod{} ranks first among discrete systems on five of six metrics, achieving a better quality--fidelity balance.
Under a matched training recipe, \OurMethod{} outperforms all compared baselines across all metric–benchmark combinations, with particularly large gains in content accuracy.
Ablations show that semantic tokens and MGM provide significant gains, while text
initialization further improves content preservation, especially under severe
degradations. 
Code and demos are available online.\footnote{https://cantabile-kwok.github.io/timuse/}

Our contributions are twofold:
{\setlength{\leftmargini}{1em}
\begin{itemize}
  \item We present \OurMethod{}, a two-stage USE framework with masked semantic token modeling that achieves a better quality–fidelity balance than previous discrete USE systems.
  \item To the best of our knowledge, we are the first to show that initialization from a pretrained text DLM benefits USE, taking a step toward DLM-based unified speech processing models.
\end{itemize}
}

\section{TIMUSE: Text-Initialized Masked Semantic Token Modeling}
\label{sec:method}

\subsection{Problem Formulation and Overview}
\label{sec:overview}

\begin{figure}[t]
  \centering
  \includegraphics[width=\linewidth]{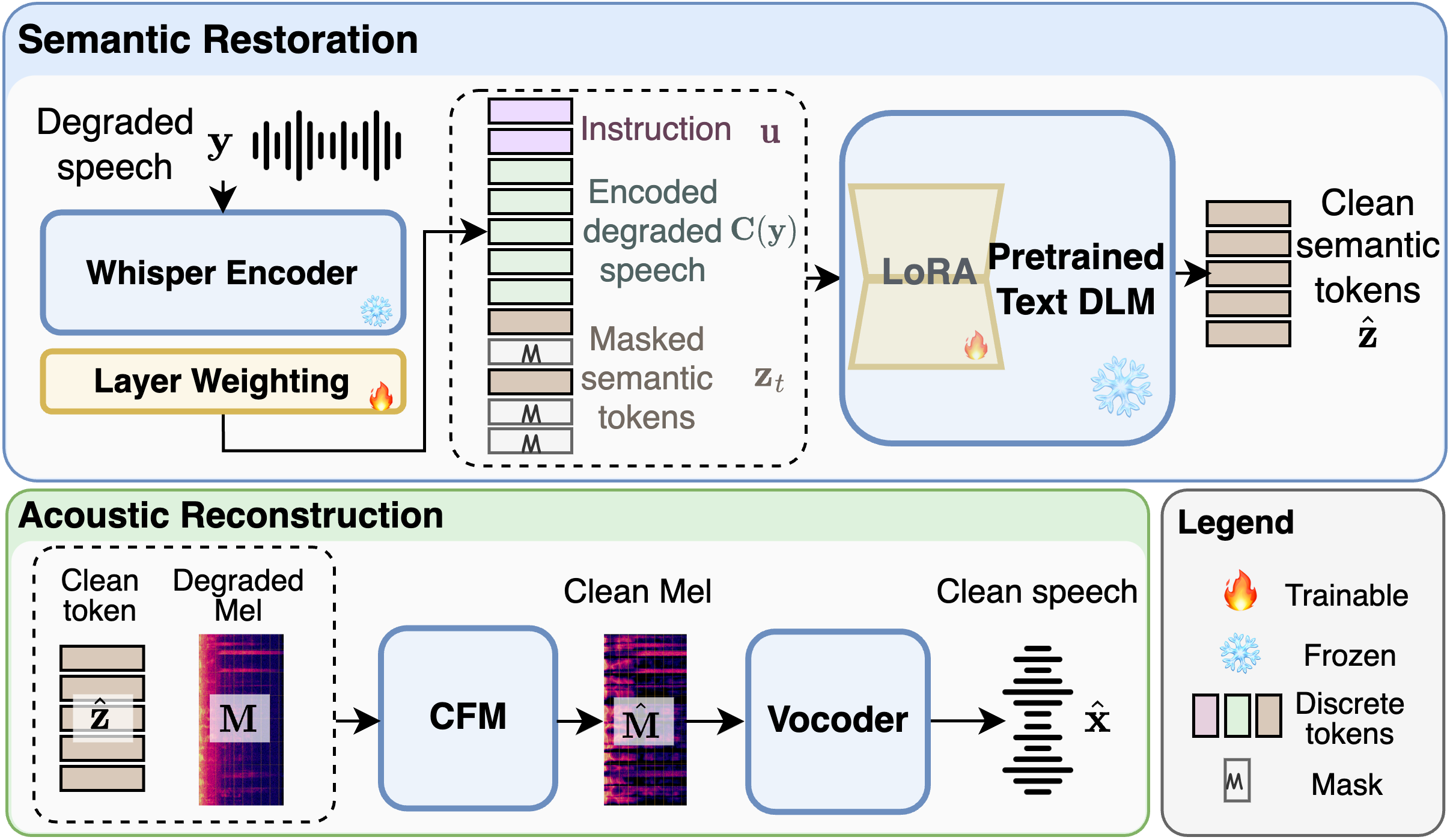}
  \caption{Overview of \OurMethod{}.}
  \label{fig:main}
\end{figure}

Let $\mathbf{y}$ and $\mathbf{x}$ denote a degraded utterance and its clean
counterpart. As illustrated in Figure~\ref{fig:main}, rather than estimating the
clean waveform directly, \OurMethod{} divides the problem into two stages.
The semantic restoration stage first predicts its clean $\mathcal{S}^3$
semantic token sequence $\hat{\mathbf{z}}$ from continuous degraded-speech
features $\mathbf{C}(\mathbf{y})$ and an optional text instruction $u$ using a
text-pretrained masked generator. The conditioning encoder $\mathbf{C}(\cdot)$
operates on 16\,kHz resampled speech.
The acoustic stage then reconstructs the
final 24\,kHz waveform from the recovered tokens while retaining the original
24\,kHz degraded waveform as an acoustic reference.

\subsection{Degraded-Speech Conditioning}
\label{sec:condition}

We implement $\mathbf{C}(\mathbf{y})$ using a frozen Whisper large-v3
encoder~\cite{radford2023whisper} applied to the degraded speech resampled to
16\,kHz. 
Rather than using the last layer output as in \cite{li2025sense}, we combine all
encoder-layer hidden states using trainable softmax-normalized scalar weights,
then apply layer normalization and a linear projection to match the hidden
dimension of the subsequent semantic restoration model. This learned mixture lets the
enhancement model perceive different acoustic and linguistic abstraction levels.

\subsection{Text-Initialized Masked Semantic Token Modeling}
\label{sec:mgm}

In the semantic restoration stage, the target is the discrete token sequence
$\mathbf{z}_0=(z_0^1,\ldots,z_0^N)$, extracted from $\mathbf x$ by a frozen
 $\mathcal{S}^3$ tokenizer~\cite{du2024cosyvoice,li2025sense} with a single codebook of
4096 entries at 50 tokens per second. 
Restoration is thus defined as predicting
this compact, explicitly clean target from continuous evidence of the degraded
observation.

We formulate the semantic restoration model as an MGM, and initialize it from a text DLM Dream-v0-Instruct-7B~\cite{ye2025dream}.
We introduce a new embedding table and output head for the $\mathcal{S}^3$ vocabulary. Its input sequence comprises an optional text instruction
$\mathbf u$, the condition
$\mathbf{C}(\mathbf{y})$, and an embedded corrupted
target sequence.
In this study, the instruction
$\mathbf u$ is randomly sampled from degradation-aware restoration templates during
training and set to a fixed generic restoration instruction
for all reported inference, leaving a natural interface for future
instruction-controlled universal speech processing tasks. The pretrained DLM backbone remains frozen and is
adapted only through LoRA~\cite{hu2022lora}; the newly introduced speech-token
embedding, output head, and condition projector are trained for the restoration
task. In this way, text pretraining supplies the
sequence-modeling initialization without substantially increasing the trainable parameter count.

We train the model with a continuous-time, discrete diffusion-style masking
process~\cite{austin2021d3pm,ye2025dream}. At each training step we sample a
diffusion timestep $t$ and convert it to a masking ratio through a bounded cosine
schedule with boundary constants $\rho_{\min},\rho_{\max}$:
\begin{equation}
  t\sim\mathcal{U}(0,1),\qquad
  \rho(t)=\rho_{\min}+(\rho_{\max}-\rho_{\min})
  \sin\!\left(\frac{\pi t}{2}\right).
  \label{eq:mask_schedule}
\end{equation}
Fully masked targets are also included during training, matching the fully
masked initial state used at inference. Given a mask set
$\mathcal{M}\sim q_{\mathrm{mask}}(\cdot\mid\rho(t))$, the corrupted target is
\begin{equation}
  z_{t}^{i}=
  \begin{cases}
    [\mathrm{MASK}], & i\in\mathcal{M},\\
    z_{0}^{i}, & \text{otherwise}.
  \end{cases}
  \label{eq:forward_corruption}
\end{equation}
The corruption operator mixes random token masking and contiguous span
masking, exposing the model to both scattered errors and
extended content loss. The sampled $t$ controls the corruption severity through
Eq.~\eqref{eq:mask_schedule}.

Conditioned on the corrupted sequence, degraded-speech features, and
instruction, the model predicts the original clean tokens. We optimize a
cross-entropy loss averaged over masked positions:
\begin{equation}
  \mathcal{L} 
  =-\mathbb{E}_{\mathbf{z}_0,t,\mathbf{z}_t}
  \left[
    \frac{1}{|\mathcal{M}|}
    \sum_{i\in\mathcal{M}}
    \log p_{\theta}\!\left(z_0^i\mid
    \mathbf{z}_t,\mathbf{C}(\mathbf{y}),\mathbf u\right)
  \right].
  \label{eq:mgm_loss}
\end{equation}
At inference, semantic generation starts from a fully masked sequence. At each
iteration, the model predicts all remaining masked positions and commits a
schedule-determined subset of the most confident predictions. Repeating this
procedure progressively produces the restored sequence $\hat{\mathbf{z}}$.

\subsection{Acoustic Reconstruction with Conditional Flow Matching}
\label{sec:cfm}

Semantic tokens retain linguistic content but omit speaker and
fine acoustic details, making an acoustic reconstruction module necessary for producing the clean speech waveform.
We formulate this module as a flow matching model in the continuous Mel domain.
We initialize it from the SenSE CFM~\cite{li2025sense} and finetune it on our training data to better match our diverse degradation conditions. 
Specifically, we extract a Mel spectrogram $\mathbf{M}$ from the original 24\,kHz degraded waveform and condition the CFM model on both this acoustic observation and the
predicted clean semantics:
\begin{equation}
  \mathbf{M}=\operatorname{Mel}(\mathbf{y}),\qquad
  \hat{\mathbf{M}}
  =\mathrm{CFM}(\hat{\mathbf{z}},
  \mathbf{M}).
  \label{eq:cfm}
\end{equation}

Unlike SenSE, during CFM sampling, we apply token-only classifier-free guidance
(CFG)~\cite{ho2022cfg}.
Let $\mathbf{m}_{\tau}$ denote the flow state at time
$\tau$. We evaluate conditional and token-null flow fields as
\begin{equation}
  \mathbf{v}_{\mathrm{c}}
  =\mathbf{v}_{\phi}(\mathbf{m}_{\tau},\tau\mid
  \hat{\mathbf{z}},\mathbf{M}),\qquad
  \mathbf{v}_{\varnothing}
  =\mathbf{v}_{\phi}(\mathbf{m}_{\tau},\tau\mid
  \varnothing,\mathbf{M}),
  \label{eq:token_only_cfg_branches}
\end{equation}
and use the guided field
$\widetilde{\mathbf{v}}=\mathbf{v}_{\mathrm{c}}
  +\gamma(\mathbf{v}_{\mathrm{c}}-\mathbf{v}_{\varnothing})$.
Unlike a fully unconditional CFG branch, the null-token branch removes only
the semantic-token condition and retains the noisy acoustic observation. CFG
therefore strengthens the contribution of restored semantics while keeping
both branches grounded in the input acoustics.
A frozen BigVGAN vocoder~\cite{lee2023bigvgan} then converts $\hat{\mathbf{M}}$
into the final 24\,kHz waveform. Here, $\hat{\mathbf{z}}$ supplies the restored clean content, while the noisy
acoustic condition provides complementary speaker and fine-grained acoustic
cues. 
We keep the CFM and vocoder fixed during semantic restoration training.

\section{Experiments}
\vspace{-0.05in}
\label{sec:experiments}

\subsection{Dataset and Training Recipe}
\vspace{-0.05in}

\subsubsection{Training Data}
We use multilingual clean speech from LibriTTS-R~\cite{koizumi2023librittsr}, the DNS5 corpus (read speech, emotional speech, and VocalSet)~\cite{dubey2023dns5}, VCTK~\cite{veaux2013vctk}, 
the URGENT 2026 training set (only SeniorTalk and EARS subsets)~\cite{li2026urgent}, ESD~\cite{zhou2021esd}, LJSpeech~\cite{ito2017ljspeech}, Expresso~\cite{nguyen2023expresso}, AISHELL-3~\cite{shi2021aishell3}, and MLS (excluding English)~\cite{pratap2020mls}.
Noise comes from the DNS5 noise set~\cite{dubey2023dns5}, FSD50K (without human voice)~\cite{fonseca2022fsd50k}, WHAM!~\cite{wichern2019wham}, DEMAND~\cite{thiemann2013demand}, and MusicCaps~\cite{agostinelli2023musiclm} as musical interference.
Room impulse responses are the simulated and real RIRs from OpenSLR SLR26 and SLR28~\cite{ko2017rir}.
This amounts to about 900 hours of clean English speech, 62 hours of clean Mandarin speech and 6k hours of other MLS languages, with 378 hours of additive noise.
During training, we first sample English, Mandarin, or other MLS languages with a fixed ratio of $6$:$1$:$2$, independent of their corpus durations, and then sample an utterance from the selected group. This substantially downweights the much larger MLS corpus while keeping English and Mandarin well represented.

\subsubsection{Training and Inference Recipe}
We train \OurMethod{} with a two-stage, 
scene-based online degradation policy. 
Each sample is drawn from a realistic scenario (identity, environmental, device, or communication), 
then assigned a difficulty-conditioned effect mixture applied on the fly. 
The degradations in these scenarios include additive noise, reverberation, bandwidth limitation, low-pass filtering, dynamic equalization, packet loss, codec reconstruction, clipping, and microphone effects.
Stage 1 emphasizes balanced everyday degradations, 
while Stage 2 increases the proportion of severe communication and multi-effect compound distortions.
Details of the simulation pipeline are provided in our open-source repository.
To better exploit contextual information, we set the maximum sample length to 10\,s.

We train \OurMethod{} with a batch size of 16 for 50k AdamW~\cite{loshchilov2019adamw} updates in Stage~1, then another 100k updates in Stage~2.
The peak learning rate is $5\times10^{-5}$ in Stage~1 and $1\times10^{-5}$ in Stage~2, with cosine annealing.
For the pretrained Dream backbone, we apply LoRA to all linear layers with rank $r=64$, scaling factor $\alpha=128$, and dropout $0.05$, with about 162M trainable parameters.
The corruption ratio follows Eq.~\eqref{eq:mask_schedule} with
$\rho_{\min}=0.15$ and $\rho_{\max}=0.95$, and $10\%$ of the steps use a fully masked input.

All semantic-token variants in the controlled comparisons and ablation studies use the same adapted CFM checkpoint, ensuring that their differences arise from semantic representation and sequence modeling rather than acoustic reconstruction.

At inference, we use two-step iterative masked decoding with a linear unmasking schedule and maximum-probability confidence, followed by eight-step CFM sampling with token-only CFG at guidance strength $\gamma=0.5$.
Under this inference setting, the average real-time factor (RTF) is 0.104 on an RTX 5090 GPU.

\subsubsection{Evaluation Data and Metrics}
We evaluate all methods on two public universal speech enhancement challenges: AATC 2025~\cite{zhang2025aatc} and URGENT 2026~\cite{li2026urgent}. 
Both benchmarks cover diverse and potentially compound degradations beyond conventional denoising and dereverberation.
The AATC 2025 test set includes 300 English and Mandarin clean-degraded pairs.
For URGENT 2026, we use the Track~1 non-blind test set with 1023 simulated pairs in multiple languages.

We evaluate USE systems with both perceptual quality and fidelity, since generative USE systems may produce perceptually plausible speech while altering linguistic content or speaker characteristics.
Perceptual metrics include DNSMOS~\cite{reddy2021dnsmos} and NISQA~\cite{mittag2021nisqa}; fidelity metrics are downstream task-derived metrics including character accuracy (CAcc, defined as $1-\text{CER}$) and speaker similarity (SpkSim) between the enhanced speech and the clean reference.
Because AATC 2025 does not provide transcripts of the clean speech, we obtain a pseudo reference by transcribing the clean speech with the same ASR model, and report pseudo-CAcc (pCAcc) on this benchmark.
The evaluation protocol follows the official URGENT 2026 implementation, with OWSM-v3.1-EBF~\cite{peng2024owsm} as the ASR model and RawNet3~\cite{jung2022rawnet3} for speaker similarity.
In short, perceptual metrics assess how natural and clean the enhanced speech sounds to humans, whereas fidelity metrics assess preservation of linguistic content and speaker identity for downstream models.

\subsection{System-Level Comparison}

\begin{table}[t]
\centering
\caption{System-level comparison on URGENT 2026. SIG, BAK, and OVRL are DNSMOS scores. Best and second-best results are in \textbf{bold} and \underline{underlined}, respectively.}
\label{tab:system}
\setlength{\tabcolsep}{2.0pt}
\resizebox{\columnwidth}{!}{%
\begin{tabular}{l cccc cc} 
\toprule
\multirow{2}{*}{\textbf{Method}}
& \multicolumn{4}{c}{\textbf{Perceptual quality}}
& \multicolumn{2}{c}{\textbf{Fidelity}} \\ 
\cmidrule(lr){2-5} \cmidrule(lr){6-7}
& \textbf{SIG$\uparrow$} & \textbf{BAK$\uparrow$} & \textbf{OVRL$\uparrow$} & \textbf{NISQA$\uparrow$}
& \textbf{CAcc$\uparrow$} & \textbf{SpkSim$\uparrow$} \\
\midrule
\multicolumn{7}{l}{\cellcolor{gray!20}\textit{Continuous-domain}} \\
BSRNN-Flow~\cite{li2025bsrnnflow} & 3.268 & 3.866 & 2.931 & 3.236 & 78.91 & \underline{0.694} \\ 
UniPASE~\cite{rong2026unipase} & 3.443 & 3.991 & 3.134 & \underline{3.952} & \textbf{86.85} & \textbf{0.720} \\ 
Sidon-v0.1~\cite{nakata2026sidon} & 3.351 & 3.987 & 3.051 & 3.377 & 76.33 & 0.507 \\ 
\midrule
\multicolumn{7}{l}{\cellcolor{gray!20}\textit{Discrete-domain}} \\
AnyEnhance~\cite{zhang2025anyenhance} & 3.162 & 3.980 & 2.877 & 3.189 & 53.51 & 0.451 \\ 
Metis-SE~\cite{wang2025metis} & \underline{3.481} & 3.978 & 3.168 & 3.901 & 74.80 & 0.511 \\ 
LLaSE-G1~\cite{kang2025llase} & 3.172 & 3.841 & 2.833 & 3.068 & 61.23 & 0.460 \\ 
UniSE~\cite{yan2026unise} & \textbf{3.508} & 3.991 & \underline{3.196} & 3.730 & 59.22 & 0.508 \\ 
SenSE~\cite{li2025sense} & 3.382 & \underline{4.054} & 3.133 & 3.707 & 80.46 & 0.592 \\ 
\OurMethod{} (Ours) & 3.477 & \textbf{4.093} & \textbf{3.225} & \textbf{4.042} & \underline{80.61} & 0.640 \\ 
\bottomrule
\end{tabular}%
}
\end{table}

\begin{table*}[t]
  \centering
  \caption{Controlled comparison on AATC 2025 and URGENT 2026. SIG, BAK, and OVRL are DNSMOS scores. pCAcc denotes pseudo character accuracy on AATC 2025. Best and second-best results in each column are in \textbf{bold} and \underline{underlined}, respectively.}
  \label{tab:controlled}
  \setlength{\tabcolsep}{3.2pt}
  \resizebox{0.90\textwidth}{!}{%
  \begin{tabular}{l cccc cc cccc cc}
  \toprule
  & \multicolumn{6}{c}{\textbf{AATC 2025}} & \multicolumn{6}{c}{\textbf{URGENT 2026}} \\
  \cmidrule(lr){2-7} \cmidrule(lr){8-13}
  \multirow{2}{*}{\textbf{Method}}
  & \multicolumn{4}{c}{\textbf{Perceptual quality}}
  & \multicolumn{2}{c}{\textbf{Fidelity}}
  & \multicolumn{4}{c}{\textbf{Perceptual quality}}
  & \multicolumn{2}{c}{\textbf{Fidelity}} \\
  \cmidrule(lr){2-5} \cmidrule(lr){6-7} \cmidrule(lr){8-11} \cmidrule(lr){12-13}
  & \textbf{SIG$\uparrow$} & \textbf{BAK$\uparrow$} & \textbf{OVRL$\uparrow$} & \textbf{NISQA$\uparrow$}
  & \textbf{pCAcc$\uparrow$} & \textbf{SpkSim$\uparrow$}
  & \textbf{SIG$\uparrow$} & \textbf{BAK$\uparrow$} & \textbf{OVRL$\uparrow$} & \textbf{NISQA$\uparrow$}
  & \textbf{CAcc$\uparrow$} & \textbf{SpkSim$\uparrow$} \\
  \midrule
  AnyEnhance & 3.331 & \underline{4.069} & 3.082 & 3.696 & 71.22 & 0.538
  & 3.287 & 3.947 & 2.990 & 3.237 & 63.31 & 0.463 \\
  UDSE & 3.169 & 4.001 & 2.924 & 3.300 & \underline{71.62} & 0.509
  & 3.036 & 3.835 & 2.738 & 2.736 & 61.71 & 0.428 \\
  SenSE & \underline{3.416} & 4.067 & \underline{3.164} & \underline{3.981} & 66.95 & \underline{0.543}
  & \underline{3.392} & \underline{4.010} & \underline{3.128} & \underline{3.860} & \underline{74.81} & \underline{0.523} \\
  \OurMethod{} (Ours)
  & \textbf{3.513} & \textbf{4.148} & \textbf{3.285} & \textbf{4.216} & \textbf{79.08} & \textbf{0.683}
  & \textbf{3.477} & \textbf{4.093} & \textbf{3.225} & \textbf{4.042} & \textbf{80.61} & \textbf{0.640} \\
  \bottomrule
  \end{tabular}}
  \end{table*}
  
  \begin{table}[t]
    \centering
    \caption{Ablation on modeling family, token type, and language-model pretraining.}
    \label{tab:ablation}
    \resizebox{\columnwidth}{!}{%
    \begin{tabular}{llc ccc}
    \toprule
    \textbf{Modeling} & \textbf{Token} & \begin{tabular}[c]{@{}c@{}}\textbf{LM}\\\textbf{Pretrain}\end{tabular} & \textbf{OVRL$\uparrow$} & \textbf{CAcc$\uparrow$} & \textbf{SpkSim$\uparrow$} \\
    \midrule
    MGM & Semantic & $\checkmark$ & 3.225 & 80.61 & 0.640 \\
    MGM & Acoustic & $\checkmark$ & 3.186 & 71.28 & 0.497 \\
    MGM & Semantic & $\times$ & 3.228 & 78.83 & 0.642 \\
    AR & Semantic & $\times$ & 3.194 & 72.84 & 0.611 \\
    \bottomrule
    \end{tabular}}
\end{table}

We first present a system-level USE comparison on URGENT 2026 using publicly released checkpoints. 
Because these systems differ in training data, model scale, simulation recipe, etc., 
this comparison only characterizes final system performance;
we leave the modeling design analysis to the controlled experiments later.
We include baselines from both the continuous and discrete domains.
Continuous-domain baselines include BSRNN-Flow~\cite{li2025bsrnnflow}, Sidon-v0.1~\cite{nakata2026sidon}, and UniPASE~\cite{rong2026unipase};
for discrete-domain models, we include AnyEnhance~\cite{zhang2025anyenhance}, LLaSE-G1~\cite{kang2025llase}, UniSE~\cite{yan2026unise}, Metis-SE~\cite{wang2025metis}, and SenSE~\cite{li2025sense}.
We use the official pretrained checkpoints for all models, specifically the AnyEnhance checkpoint\footnote{https://github.com/viewfinder-annn/AnyEnhance-v1.} and the URGENT 2026 baseline for BSRNN-Flow\footnote{https://github.com/urgent-challenge/urgent2026\_challenge\_track1}.

As shown in Table~\ref{tab:system}, \OurMethod{} ranks first among discrete-domain systems on five of the six reported metrics. 
Relative to the best competing discrete-domain result for each metric, it improves BAK by 0.039, OVRL by 0.029, NISQA by 0.141, CAcc by 0.15 percentage points, and SpkSim by 0.048. Its BAK, OVRL, and NISQA are also the best among all evaluated systems. Although UniSE and Metis-SE obtain slightly higher SIG scores, \OurMethod{} offers the strongest overall balance among discrete systems. The controlled studies below examine the sources of these gains.

\subsection{Controlled Experiments}
\label{sec:controlled}

We next compare models under a matched training dataset and recipe to eliminate the effects of training-data size, quality, and simulation strategy.
We consider three discrete-domain USE methods: AnyEnhance, UDSE~\cite{liu2025udse}, and SenSE.
We adapt their official training code to our online two-stage recipe and train all models for 150k updates with the same batch size.

Table~\ref{tab:controlled} shows that \OurMethod{} ranks first in all twelve metric--benchmark combinations.
Beyond CAcc gains of 7.46 percentage points on AATC 2025 and 5.80 points on URGENT 2026, improvements in perceptual quality and speaker similarity show that stronger content recovery does not sacrifice speaker identity.
Because all systems share the training data and recipe, these matched gains more directly reflect the model design; the following ablations separate the contribution of each choice.

\vspace{-0.1in}
\subsection{Ablation and Analysis}

We further isolate the effect of three core design choices: generative modeling family (MGM vs.\ AR), discrete token family (semantic vs.\ acoustic), and text LM pretraining.
To isolate the token family, we replace the single-codebook $\mathcal{S}^3$ tokenizer with the likewise single-codebook WavTokenizer-large~\cite{ji2025wavtokenizer} at 24\,kHz.
For all no-pretrain variants, including both semantic- and acoustic-token MGM, we use a randomly initialized Transformer with approximately 162M trainable parameters, matching the trainable budget of our LoRA adapter on Dream DLM.
To compare MGM and AR, we hold the Transformer architecture fixed and change only the sequence objective and decoding procedure.
These comparisons are even stricter than those in Section~\ref{sec:controlled}, since other variables such as trainable parameter budget and audio encoder are held fixed.

Table~\ref{tab:ablation} reveals a clear hierarchy among the three design choices. Replacing semantic tokens with acoustic tokens causes the largest loss, particularly in content and speaker fidelity, establishing the semantic target as a primary source of the advantage. With the token type and parameter budget fixed, MGM also consistently outperforms AR; the final AR row closely resembles SenSE, i.e., autoregressive semantic-token prediction without text-LM initialization. Initializing MGM from the text-pretrained DLM then provides a complementary gain: CAcc increases by 1.78pp, while OVRL and SpkSim remain essentially unchanged. 
Thus, semantic targets and masked modeling drive the overall improvement, whereas text initialization more specifically strengthens content preservation.

\begin{figure}[t]
  \centering
  \includegraphics[width=\columnwidth]{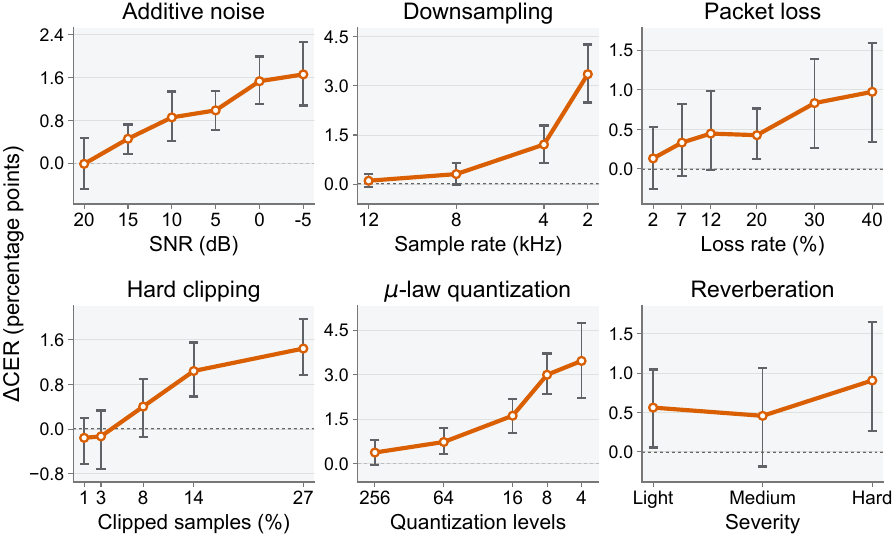}
\vspace{-0.2in}
  \caption{CER gains of text initialization across degradations and severities. Error bars denote 95\% confidence intervals.}
  \label{fig:severity}
\end{figure}

To examine when this targeted gain arises, we create severity-controlled test sets by applying one degradation at a time to the clean reference utterances of URGENT 2026. We compare \OurMethod{} with the parameter-matched, randomly initialized semantic-token MGM in the third row of Table~\ref{tab:ablation}, using the same corrupted utterances for both systems. Figure~\ref{fig:severity} reports the corpus-level CER of this model minus that of \OurMethod{}, in percentage points; positive values therefore favor text initialization. The pointwise 95\% confidence intervals are obtained by paired bootstrap resampling of utterances. The gain generally increases as noise, downsampling, clipping, and quantization become more severe, and is also positive at high packet-loss rates. Several mild conditions remain indistinguishable. 
These trends support our central hypothesis that text-pretrained linguistic priors become increasingly valuable as acoustic evidence is lost, establishing text DLM initialization as an effective mechanism for content recovery under severe degradations.

\vspace{-0.05in}
\section{Conclusion}

\label{sec:conclusion}

We presented \OurMethod{}, a two-stage USE framework combining text-initialized
masked semantic-token prediction with flow-matching-based acoustic
reconstruction. On URGENT 2026, it offers a strong overall balance among
discrete systems, while under matched training it leads all metrics on AATC 2025
and URGENT 2026. Ablations show that semantic token targets and masked modeling drive
the overall gains, whereas initialization from a text DLM selectively improves content
preservation, particularly under severe content-degrading conditions. 
These results support the use of text-pretrained priors for USE and highlight the potential of DLMs for future universal speech signal processing.
\clearpage

\section{Compliance with Ethical Standards}
\vspace{-0.1in}
This study used publicly available speech data; no additional ethical approval was required.

\vspace{-0.2in}
\bibliographystyle{IEEEbib}
\let\defaultthebibliography\thebibliography
\renewcommand{\thebibliography}[1]{%
  \defaultthebibliography{#1}%
  \setlength{\itemsep}{-0.3ex}%
}
\bibliography{references}

\end{document}